\documentstyle[aps,epsf]{revtex}  

\newcommand{\simleq}{\, ^<_{\sim} \,}
\begin{document}        

\baselineskip 14pt
\title{Deep-Inelastic $e^+p$ Scattering at Very High $Q^2$ 
       from ZEUS at HERA }
\author{C.M.~Ginsburg}
\address{The Ohio State University, 
         Physics Department, 
         Columbus, OH 43210-1106, USA \\ 
         on behalf of the ZEUS Collaboration}

\maketitle

\begin{abstract}        
     Neutral-current and charged-current deep-inelastic scattering
at very high four-momentum transfer squared ($Q^2$) have been studied in 
positron-proton collisions at center-of-mass energy 300 GeV 
using the ZEUS detector at HERA.  An integrated luminosity of 
47.7 pb$^{-1}$ was collected in the years 1994-1997.
Differential cross sections are presented for $Q^2>400$ GeV$^2$
and compared to Standard Model predictions.
\end{abstract}          

\section{Deep-Inelastic scattering at HERA}
Deep-inelastic scattering (DIS) provides a wealth of information about
nucleon structure.
Recently, very high momentum transfers in DIS have been achieved at the
HERA collider,
where 820 GeV protons have been collided with 27.5 GeV positrons for 
a center-of-mass energy $\sqrt{s}$=300 GeV.
In the highest momentum transfer region, $e^+p$ DIS cross sections depend 
on proton parton densities and properties of the electroweak interaction.
 
\begin{figure}
\centerline{\hspace*{-1.cm}
   \epsfxsize 16 truecm \epsfbox{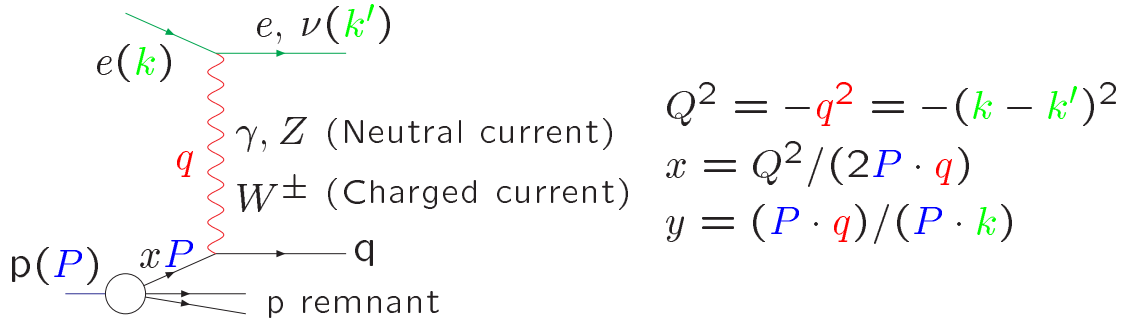}}   
\caption{The $e^{\pm}p$ deep-inelastic scattering process.
 \label{fig:dis}}
\end{figure}

The $e^{\pm}p$ DIS process is illustrated in Fig.~\ref{fig:dis}.
The variables used to describe the process are
$x$, the struck parton momentum fraction,
$y$, the fractional energy transfer in the proton rest frame (inelasticity)
and $Q^2$, the four-momentum transfer squared, where $Q^2 = sxy$.
Neutral-current DIS events are characterized by the exchange of
a photon or $Z$-boson, and have a positron and a jet (or jets) 
in the final state.  The outgoing positron and the hadronic matter
are balanced in transverse momentum.
Charged-current DIS events are characterized by the exchange
of a $W$-boson, and contain an undetected neutrino 
and a jet (or jets) in the final state.  The presence of the neutrino 
is detected as missing transverse momentum.

The DIS data presented here were collected and analyzed by the ZEUS 
collaboration and correspond to an integrated luminosity of 
47.7 pb$^{-1}$  taken from 1994 to 1997.

\section{The ZEUS Detector}
\begin{figure}
\centerline{\hspace*{-0.5cm}
            \epsfxsize 13 truecm \epsfbox{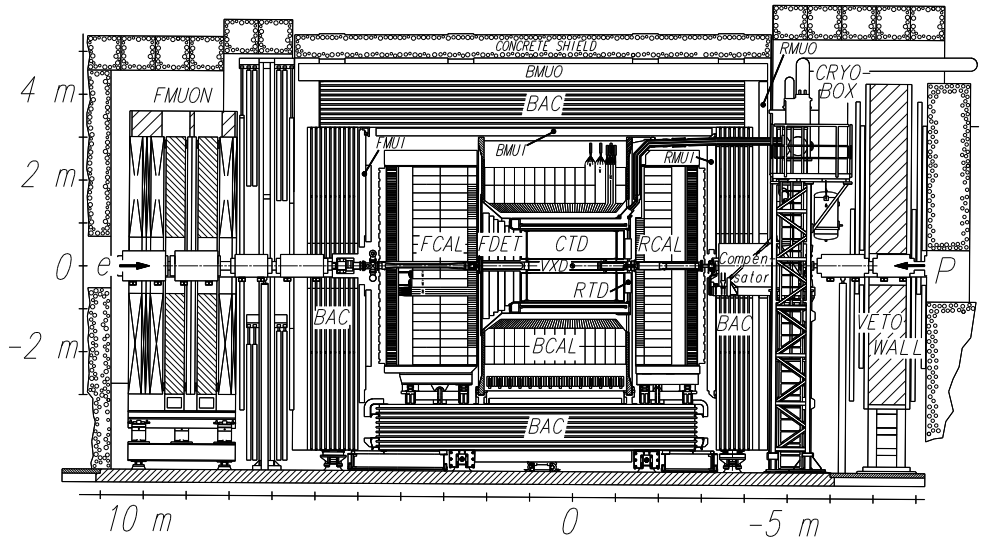}}
\caption{The ZEUS detector. 
 \label{fig:detector}}
\end{figure}

ZEUS~\cite{ref:zeusdet} is a  multi-purpose magnetic detector;
the primary components used in these analyses are the calorimeters
(RCAL,BCAL,FCAL), 
the central tracking detector (CTD), and the luminosity monitor.
The coordinate system is defined such that
the $z$-axis follows the proton direction, and the origin is
the nominal $ep$ interaction point.  
The ZEUS detector is displayed in Fig.~\ref{fig:detector}.
 
The ZEUS compensating uranium-scintillator calorimeter covers the polar 
angle region 
$2.6^{\circ} < \theta < 176.1^{\circ}$ with full azimuthal coverage
over this region.
Its energy resolution for electromagnetic showers is 
$\sigma_E/E \simeq 18 \% / \sqrt{E(\rm{GeV})}$, and 
for hadronic showers is
$\sigma_E/E \simeq 35 \% / \sqrt{E(\rm{GeV})}$, as measured under
test-beam conditions.

The ZEUS CTD operates in a solenoidal
1.43 T magnetic field,
and primarily provides vertex reconstruction,
track momentum, and charge information for these analyses.

The luminosity is determined from the rate of Bethe-Heitler bremsstrahlung 
($ep\rightarrow ep\gamma$) photons detected in 
an electromagnetic calorimeter at $z=-107$~m.

\section{Structure Function Formalism}
To lowest order (QED Born level) the neutral-current DIS
($e^+p \rightarrow e^+X$) cross section is
\begin{equation}
  \frac{d^2\sigma^{NC}(e^+)}{dx\,dQ^2}=\frac{2\pi\alpha^2}{xQ^4}\,
  \left[Y_{\small +} \, {\cal F}_2^{NC}(x,Q^2) 
        - Y_{\small -} \, x{\cal F}_3^{NC}(x,Q^2)
        - y^2 \,  {\cal F}_L^{NC}(x,Q^2) \strut\right] 
\end{equation}
where $Y_{\small \pm} = 1 \pm (1-y)^2$.
In lowest-order QCD, the structure functions ${\cal F}_2^{NC}$ and 
$x{\cal F}_3^{NC}$ are the sums over quark flavor of the product of 
quark couplings and momentum distributions.  The quark couplings
depend on the quark charges, and the electroweak parameters 
$M_Z, {\rm sin}^2\theta_W$, etc.

The QED Born-level charged-current DIS ($e^+p \rightarrow \bar{\nu}_e X$) 
cross section is
\begin{equation}
\frac{d^2\sigma^{CC}(e^+)}{dx\,dQ^2}= 
      \frac{G_F^2}{4\pi x} \! 
      \left( \frac{M_W^2}{Q^2+M_W^2} \right)^2 \! \!
      \left[Y_{\tiny +} \, {\cal F}_2^{CC}(x,Q^2) 
          - Y_{\tiny -} \, x{\cal F}_3^{CC}(x,Q^2)
         - y^2 {\cal F}_L^{CC}(x,Q^2) \strut  \right]
\end{equation}
where at lowest order the structure functions ${\cal F}_2^{CC}$ 
and  $x{\cal F}_3^{CC}$
contain sums and differences of quark and antiquark momentum distributions.

The neutral- and charged-current longitudinal 
structure functions, ${\cal F}^{NC}_L$ and ${\cal F}^{CC}_L$, respectively,
provide a small ($\sim 1\%$) contribution in the 
kinematic range discussed here, and have been included.

Electroweak radiative corrections to these Born-level equations, 
including initial- and final-state radiation, vertex and propagator 
corrections, and two-boson exchange,
are significant and have been included to at least lowest 
order~\cite{ref:radcorr}.

\section{The Very High-$Q^2$ Neutral-Current and Charged-Current Data }
\begin{figure}
\centerline{\hspace*{-0.15cm} 
                  \epsfysize 5.1 truecm \epsfbox{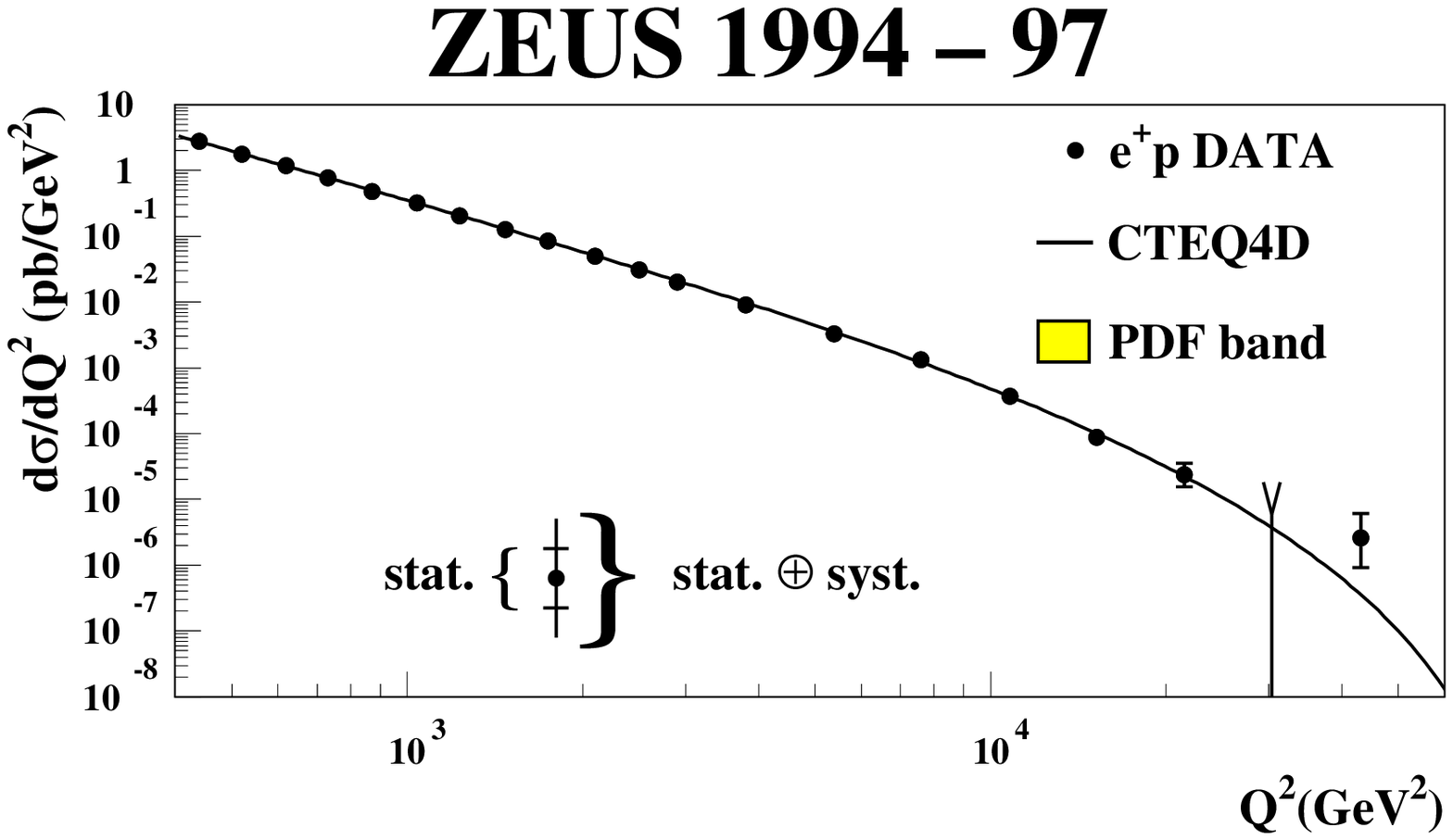} 
  \hspace*{0.15cm} \epsfysize 5.1 truecm \epsfbox{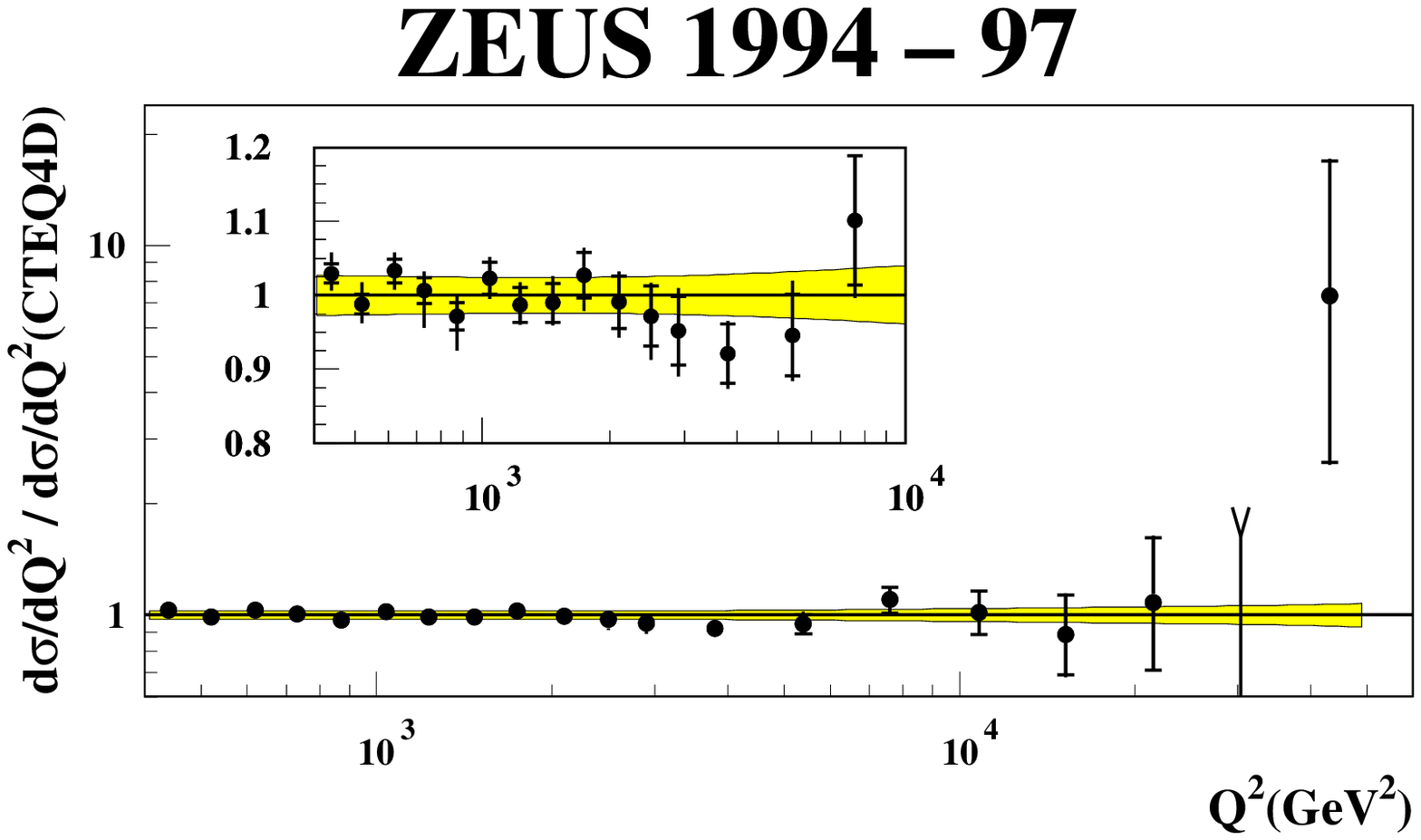} }
\caption{The neutral-current DIS (a) differential 
  cross section  $d\sigma /dQ^2$,
  and (b) ratio of data to SM prediction with blow-up view of
$400 < Q^2 < 10000$ GeV$^2$ region (inlaid).  
 \label{fig:nc_dsdq2}}
\end{figure}

\begin{figure}
\centerline{\epsfysize 9.7 truecm \epsfbox{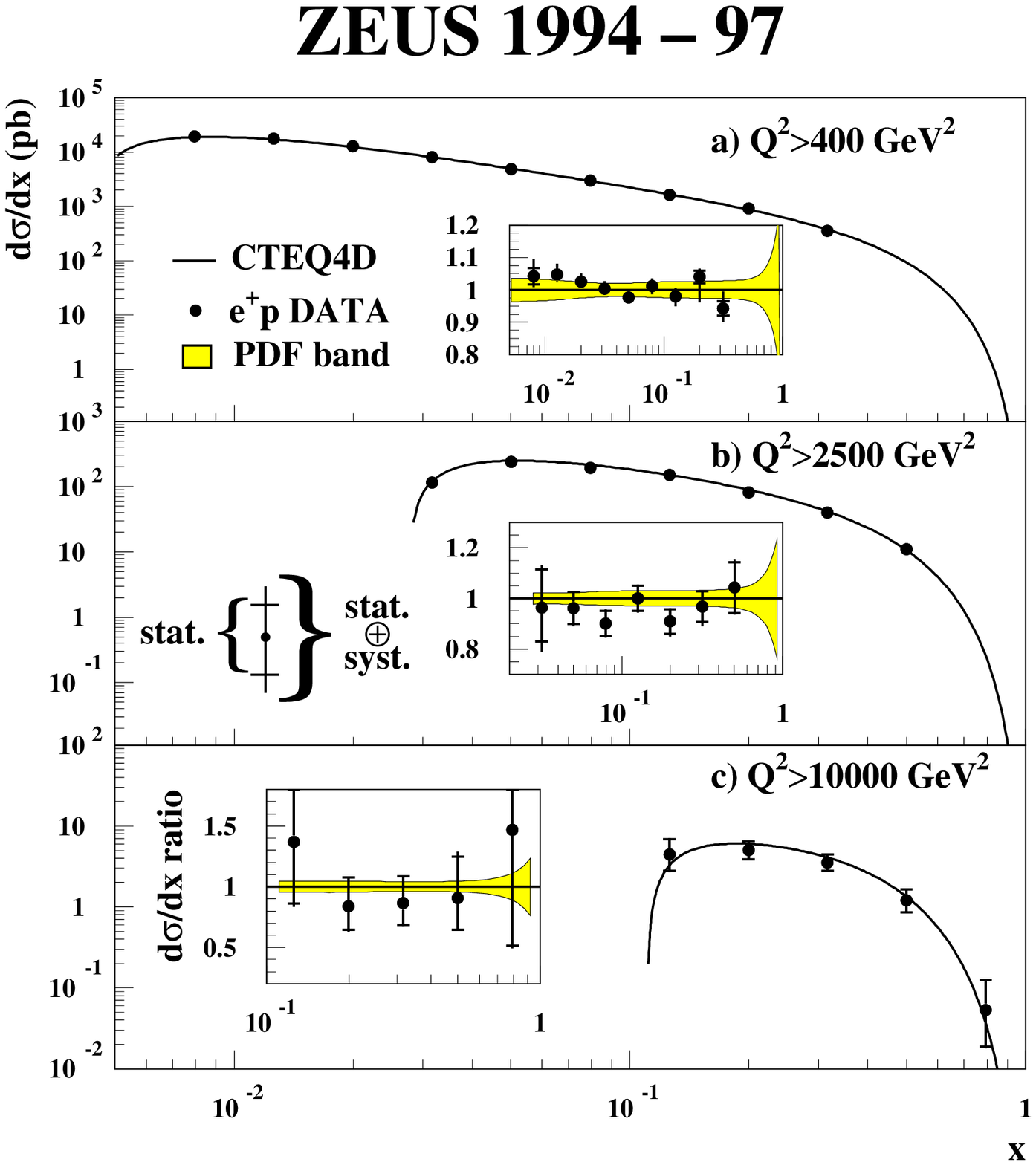}
            \hspace*{0.4 truecm}
            \epsfysize 9.7 truecm \epsfbox{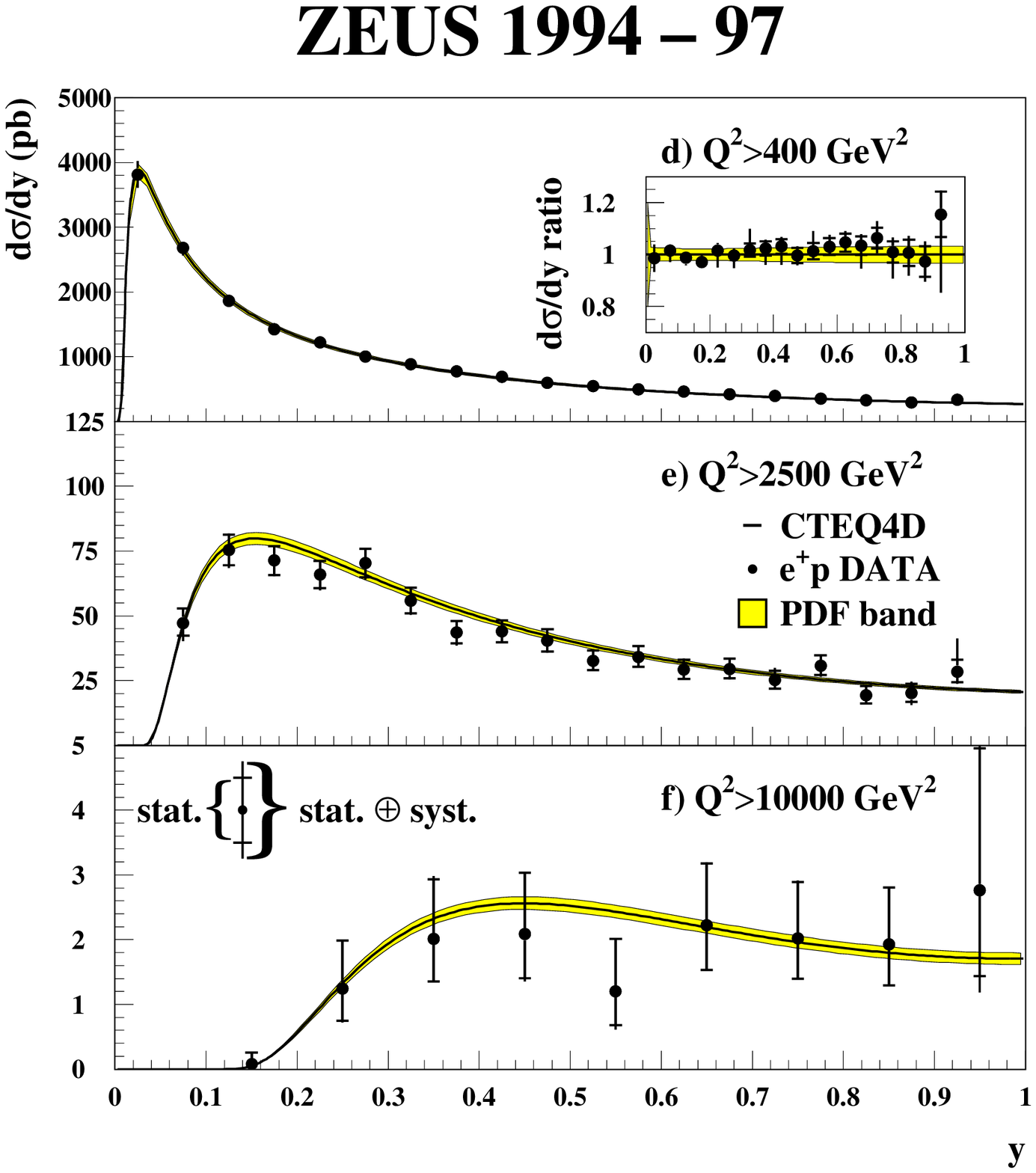} }
\caption{The neutral-current DIS differential 
     cross sections $d\sigma / dx$ (a-c) and $d\sigma / dy$ (d-f) 
for three different minimum-$Q^2$ bins: 
(a,d) $Q^2 > 400$ GeV, (b,e) $Q^2 > 2500$ GeV, and (c,f) $Q^2 > 10000$ GeV.  
The inlaid plots show the ratio of the data
to the Standard Model prediction.
 \label{fig:nc_dsdx_dsdy}}
\end{figure}

\begin{figure}
\centerline{\epsfysize 6.7 truecm \epsfbox{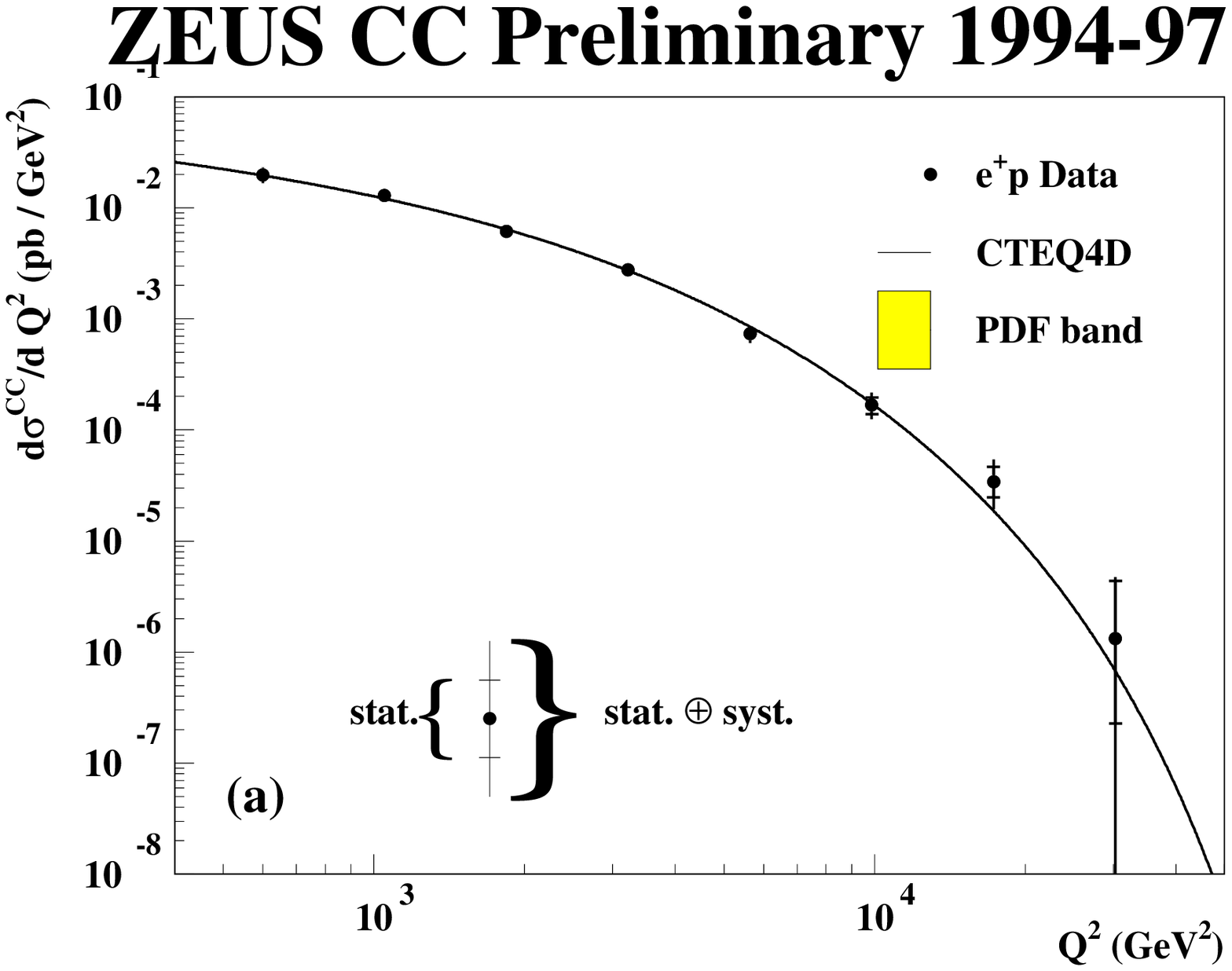} 
  \hspace*{0.5cm} \epsfysize 6.7 truecm \epsfbox{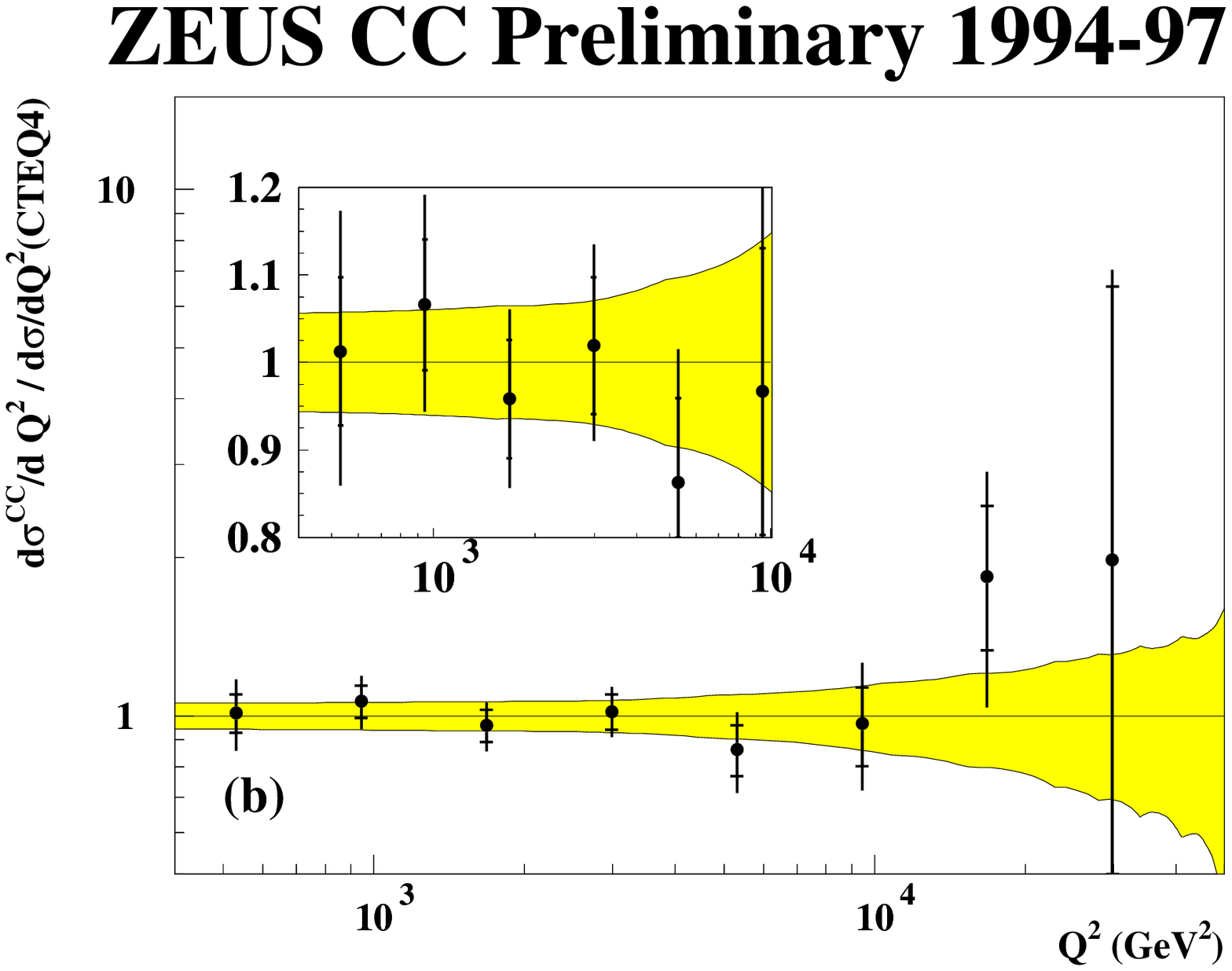} }
\caption{The charged-current DIS (a) differential 
  cross section  $d\sigma /dQ^2$,
and (b) ratio of data to SM prediction with blow-up view of
$400 < Q^2 < 10000$ GeV$^2$ region (inlaid).  
 \label{fig:cc_dsdq2}}
\end{figure}

\begin{figure}
\centerline{\epsfysize 6.4 truecm \epsfbox{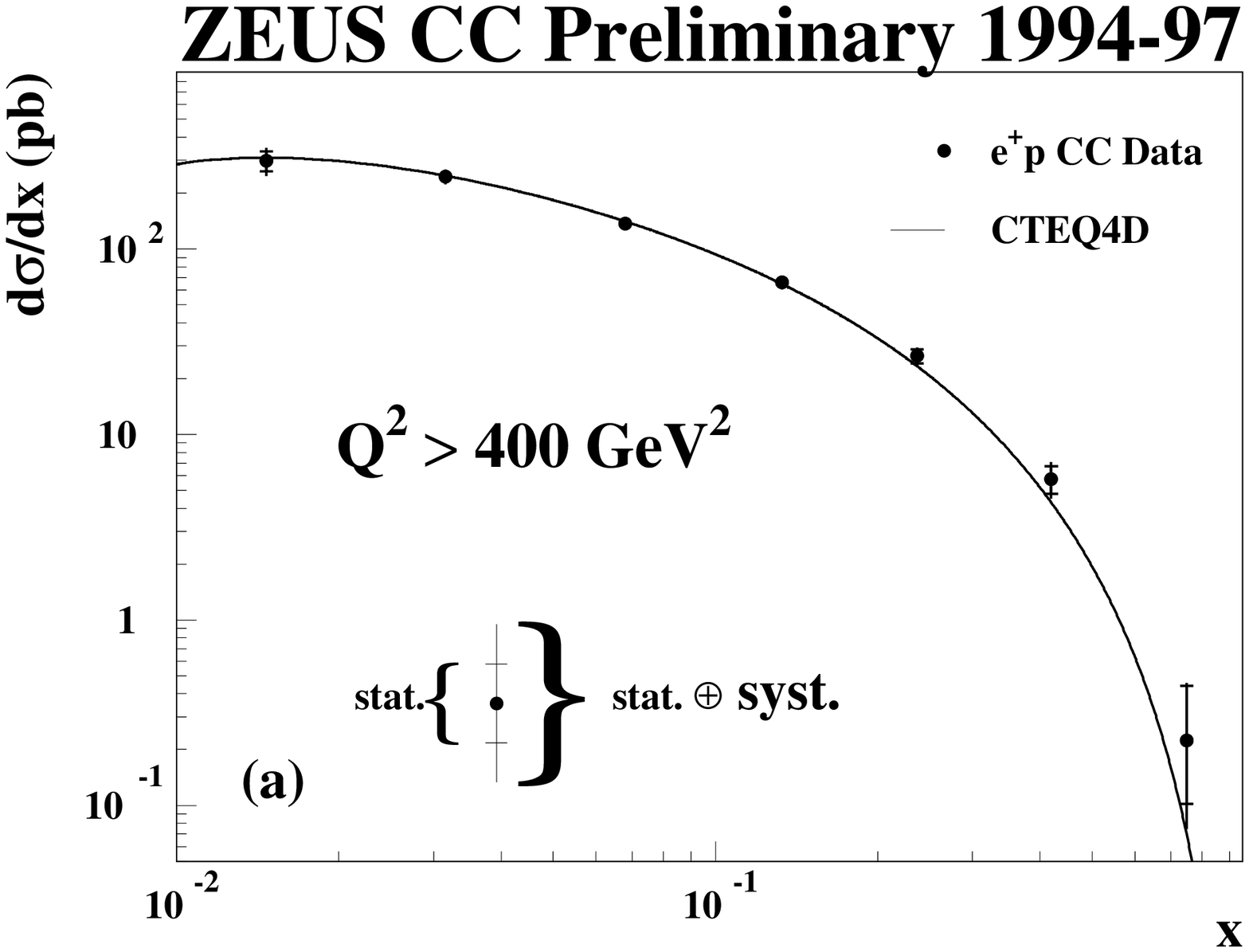} 
            \epsfysize 6.4 truecm \epsfbox{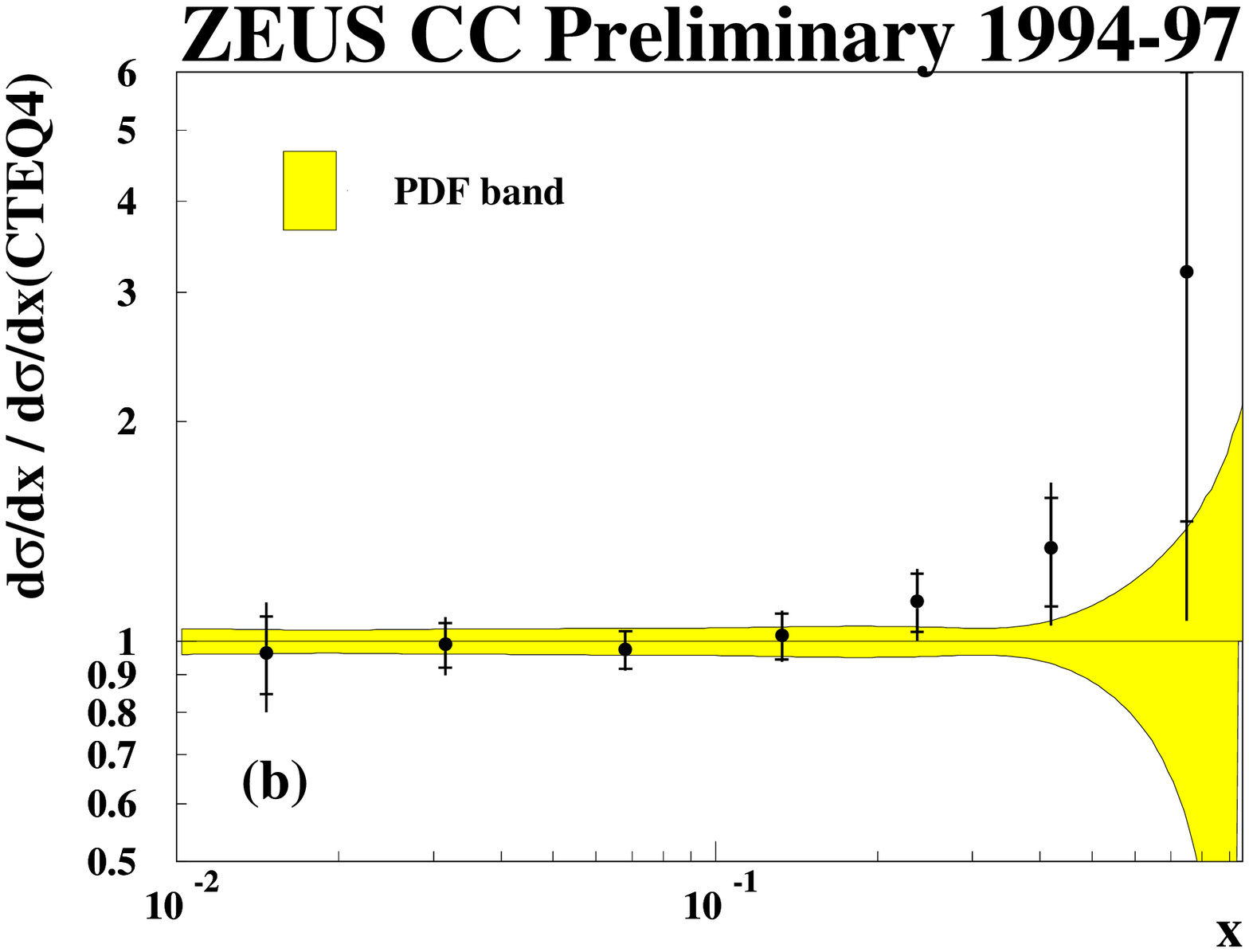} }
\caption{The charged-current DIS (a) differential cross section
$d\sigma / dx$, 
(b) ratio of data to SM prediction.
 \label{fig:cc_dsdx}}
\end{figure}

\begin{figure}
\centerline{\epsfysize 6.4 truecm \epsfbox{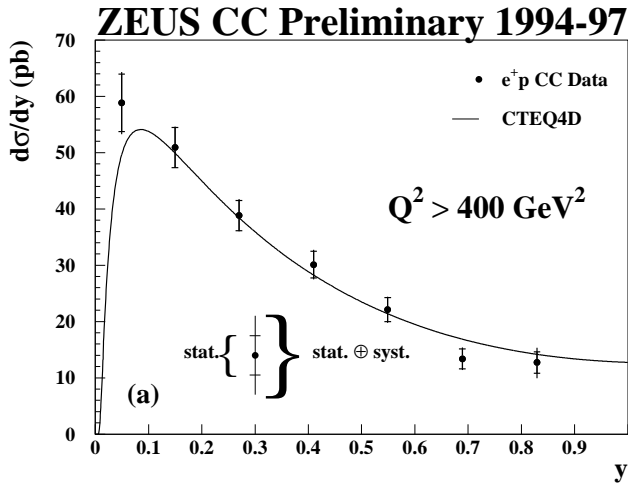} 
            \epsfysize 6.4 truecm \epsfbox{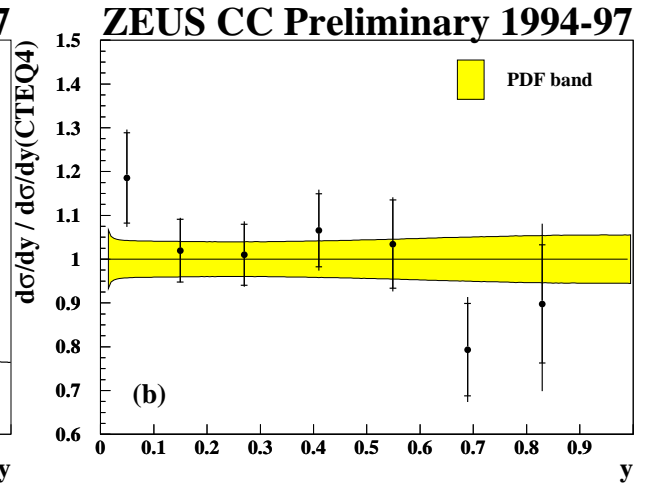} }
\caption{The charged-current DIS (a) differential cross section 
$d\sigma / dy$,
and (b) ratio of data to SM prediction.
 \label{fig:cc_dsdy}}
\end{figure}

The differential NC DIS cross section
$d\sigma^{NC}/dQ^2$ is shown in Fig.~\ref{fig:nc_dsdq2};
$d\sigma^{NC}/dx$ and $d\sigma^{NC}/dy$ are shown 
in Fig.~\ref{fig:nc_dsdx_dsdy}.
The differential CC DIS cross sections
$d\sigma^{CC}/dQ^2$, $d\sigma^{CC}/dx$, and $d\sigma^{CC}/dy$, are shown 
in Figs.~\ref{fig:cc_dsdq2}, ~\ref{fig:cc_dsdx}.
and~\ref{fig:cc_dsdy}, respectively.
For both neutral and charged current, the data points are compared
to the Standard Model predictions using
the CTEQ4D~\cite{ref:cteq} parton distribution functions (PDF's), 
shown by the solid curves, with estimated PDF uncertainty 
shown by the shaded bands.


The PDF uncertainty is calculated from a NLO fit~\cite{ref:qcdfits} to world
DIS data, and includes statistical and systematic errors on these
data, as well as variations in the assumed electroweak and QCD parameters.
For the neutral-current cross section, these uncertainties range 
from 2.5\% at $Q^2=400 \ {\rm GeV}^2$ to 8\% at $Q^2=40000 \ {\rm GeV}^2$. 
For the charged-current cross section, the extracted uncertainties range 
from 9\% at $Q^2=400 \ {\rm GeV}^2$ to 17\% at $Q^2=10000 \ {\rm GeV}^2$. 
The larger CC uncertainty is due to the larger uncertainty in the $d$-quark 
PDF relative to the $u$-quark PDF.  
Both the NLO fit, which includes higher-twist effects, and a recent 
reanalysis of NMC and SLAC data~\cite{ref:bodekyang} yield a larger 
$d/u$ ratio at high-$x$ than the CTEQ4 PDF's, where the $d/u$ ratio
is constrained to be zero at $x=1$.  Within present experimental precision,
any of these hypotheses can be accommodated;
their differences are not included in the PDF uncertainty band.  
Increasing the $d/u$ ratio at high-$x$ reduces the CC data excess
at high-$x$, but does not appreciably affect the NC cross section
since the NC process is not sensitive to the $d$-quark.

\begin{figure}
\centerline{\epsfysize 12 truecm \epsfbox{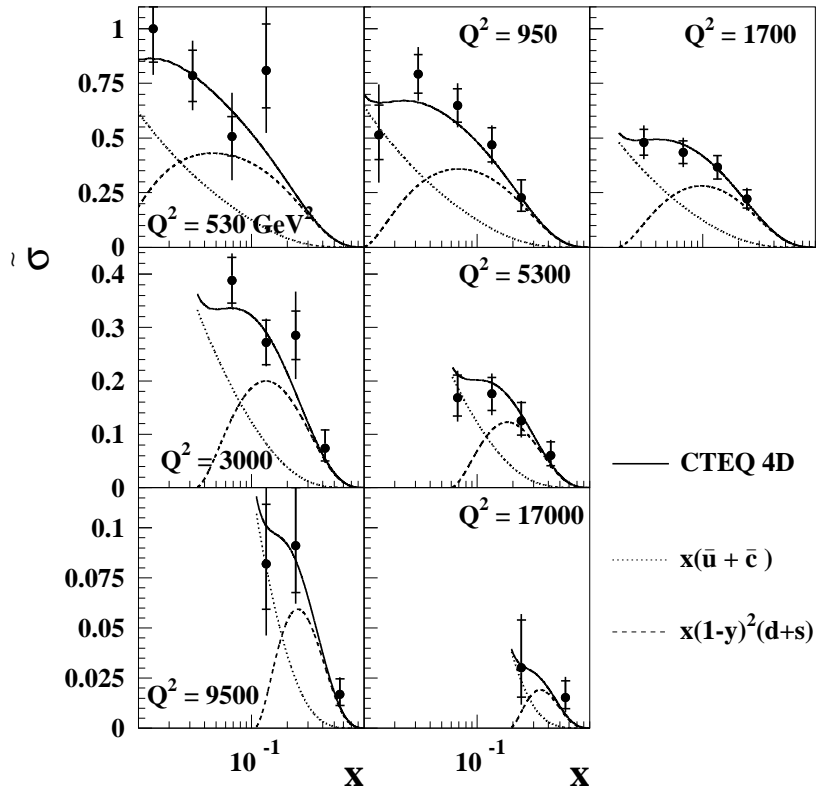} }
\caption{The charged-current DIS reduced cross section data as a function
of $x$ in bins of $Q^2$, compared to the Standard Model predictions.
 \label{fig:ccreduced}}
\end{figure}

The charged-current DIS reduced cross section
\begin{equation}
\tilde{\sigma} = \frac{d^2\sigma}{dxdQ^2} \cdot \frac{2\pi x}{G_F^2} \cdot
     \left( \frac{M_W^2}{Q^2+M_W^2} \right)^{-2} \\
= x(\bar{u} + \bar{c}) + x(1-y)^2(d+s) 
\end{equation}
is shown in Fig.~\ref{fig:ccreduced} along with the
Standard Model (CTEQ4D) prediction.
At high-$x$, the valence 
$d$ and $s$ quarks (dashed curves) dominate $\tilde{\sigma}$, 
whereas at lower-$x$ the 
$\bar{u}$ and $\bar{c}$ sea quarks (dotted curves) dominate.

At momentum-transfer-squared close to the
$Z$ and $W$ masses squared, i.e., $Q^2 {\, ^>_{\sim} \,} 6000-8000$ GeV$^2$, 
the cross sections become sensitive to contributions from these exchanges,
and to the propagator masses in particular.  

\begin{figure}
\centerline{\epsfysize 5.0 truecm \epsfbox{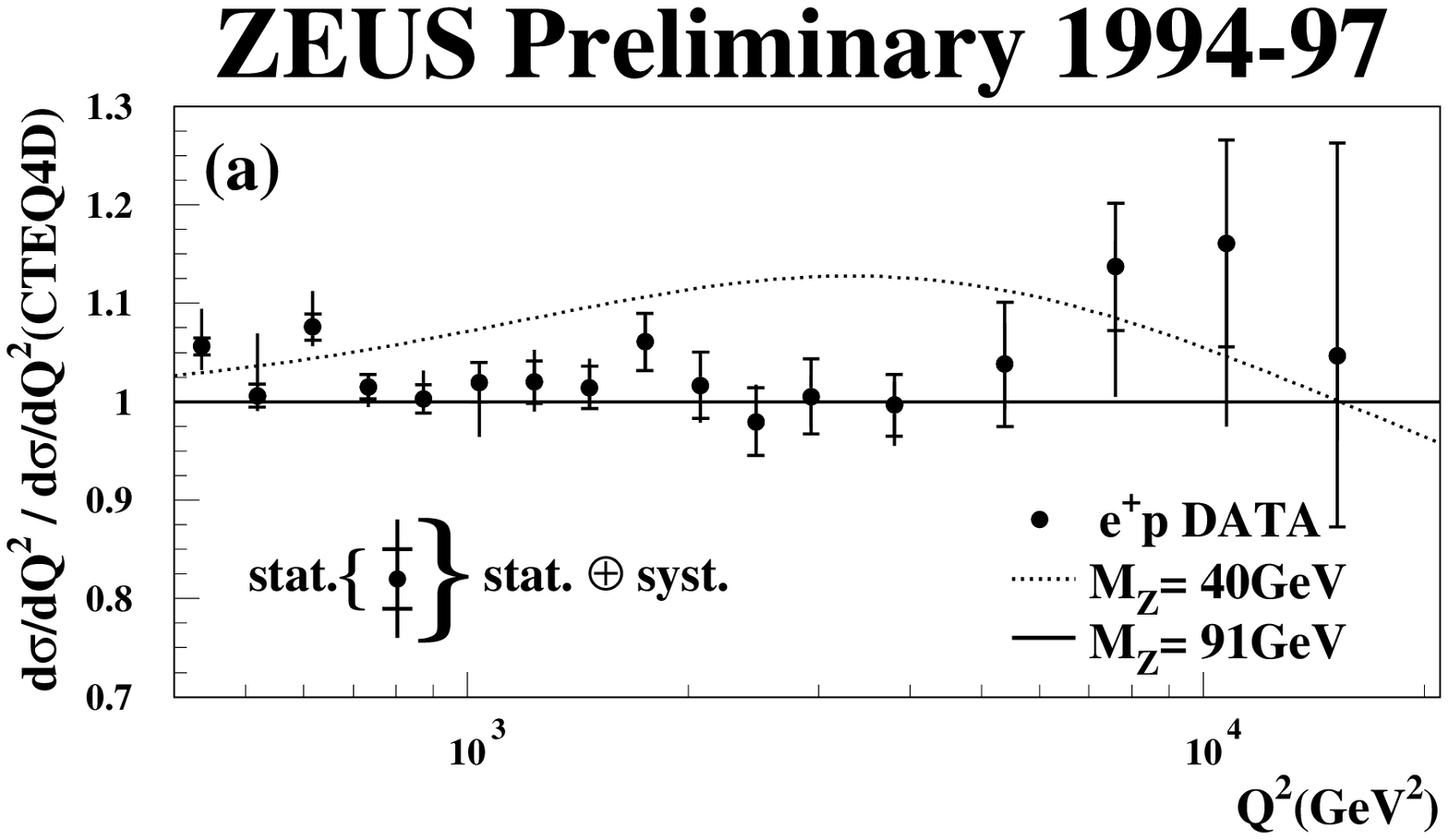} 
            \epsfysize 5.0 truecm \epsfbox{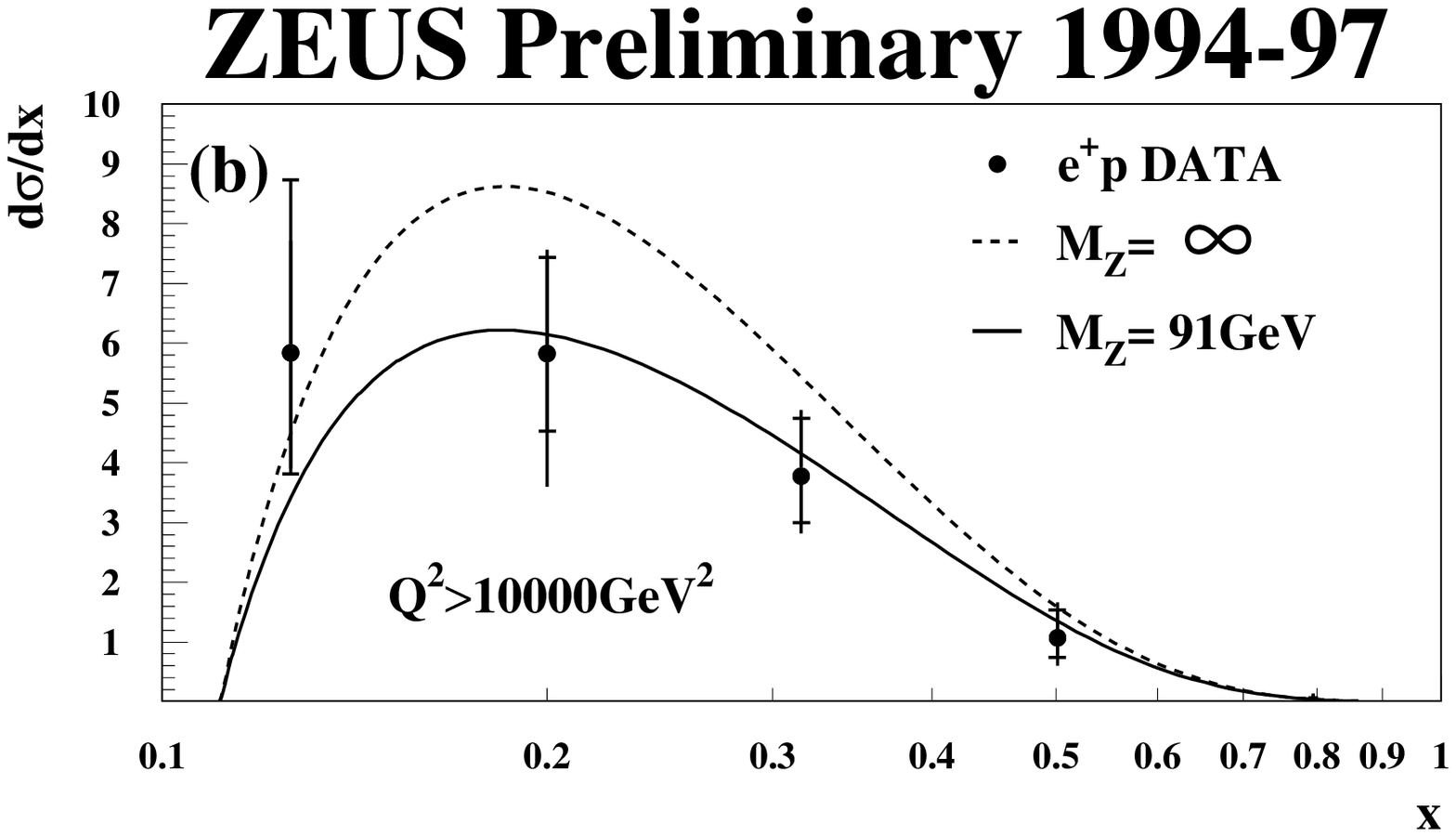} }
\caption{Sensitivity of the NC cross section to the $Z$ mass: 
(a) the ratio of NC differential cross section $d\sigma/dQ^2$ to the 
SM prediction (solid line), 
and with the assumption of $M_Z=40$ GeV (dashed curve); and (b)
the $Q^2>10000$ GeV$^2$ NC differential cross section $d\sigma/dx$
compared to the SM prediction (solid curve), and with the assumption
of an infinite $Z$ mass (dashed curve).
 \label{fig:zmass}}
\end{figure}

The sensitivity of the NC cross section to the $Z$ mass is demonstrated
in Fig.~\ref{fig:zmass}.  The measured cross sections
are compared with the Standard Model predictions by varying $M_Z$ 
while keeping the couplings fixed.
Three mass values are considered, $M_Z$ = 40, 91 and $\infty$ GeV.
Clearly, $M_Z \sim 91$ GeV is favored, in agreement with the
world average value $M_Z$ = (91.187 $\pm$ 0.007) GeV~\cite{ref:pdb98}.

\begin{figure}
\centerline{\epsfxsize 10 truecm \epsfbox{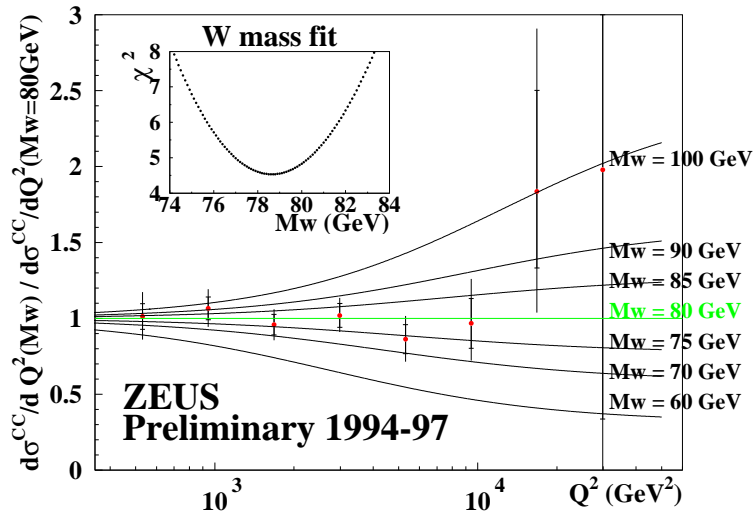} }
\caption{Sensitivity of the CC cross section to the $W$ mass, shown
as a ratio of $d\sigma/dQ^2$ to the SM prediction, for several values
of $M_W$ between 60 and 100 GeV.  The inlaid plot shows the $\chi^2$
of the $M_W$ fit.
 \label{fig:wmass}}
\end{figure}

In the CC channel, the sensitivity of the cross section
to the $W$ mass is better than in the NC case, as shown in
Fig.~\ref{fig:wmass}.  The $M_W$ may be extracted by
a $\chi^2$ fit to the measured cross sections, 
leaving all other electroweak parameters fixed, with the result
\begin{equation}
M_W = 78.6^{+2.5}_{-2.4} ({\rm stat.})^{+3.3}_{-3.1} ({\rm syst.}) \ {\rm GeV} .
\end{equation}
This may be compared with the world average
value of $M_W$ = (80.41 $\pm$ 0.10) GeV~\cite{ref:pdb98}.

The agreement between these $Z$ and $W$ masses and the
masses extracted through timelike production in $e^+e^-$ and $\bar{p}p$ data
confirms the Standard Model prediction in the spacelike regime.


\section{Conclusions}
The ZEUS neutral-current and charged-current DIS cross sections
in the momentum transfer range $400 < Q^2 \simleq 50000$ GeV$^2$, 
i.e., for $Q^2$ over more than two orders of magnitude,
have been shown to be in good agreement with the Standard Model 
predictions.  However, a slight excess persists at highest $Q^2$.  
With the current data set, ZEUS has gained sensitivity
to the electroweak propagator masses $M_W$ and $M_Z$.

\section{Outlook}
In 1998, HERA switched from positron-proton collisions to 
electron-proton collisions.  
In addition, the proton energy was increased from 820 GeV
to 920 GeV, thus increasing the center-of-mass energy by 6\% 
from 300 GeV to 318 GeV.
As of January 1, 1999, ZEUS had already accumulated $4.5 {\rm pb}^{-1}$
of $e^-p$ data, nearly an order of magnitude more than accumulated
during the only other $e^-p$ running period 1992-1993.  
An additional $\sim 35 {\rm pb}^{-1}$ is expected in the 1999 
calendar year.  
The HERA upgrade begins in May 2000, after which we foresee
$\sim 150 {\rm pb}^{-1}$/year.  
These future data will improve our understanding of the electroweak
interaction, and may even hold some surprises.
We anticipate an exciting future for high-$Q^2$ physics at HERA.
\begin{acknowledgements}
The kind assistance of my ZEUS colleagues in the preparation of
this talk, and the great efforts of the conference organizers
are greatly appreciated.  This work is partly funded by
the U.S. Department of Energy.
\end{acknowledgements}


\end{document}